\documentclass[
floatfix,
aps, 
prb,
amsmath,
twocolumn,
superscriptaddress,
amssymb,
tightenlines,
]{revtex4-1}

\pdfoutput=1
\usepackage{amsfonts,amssymb}
\usepackage[subnum]{cases}
\usepackage{mathrsfs}
\usepackage{amsmath}
\usepackage[none]{hyphenat}
\usepackage{hyperref} % links
\usepackage{bm} % bold math letters
\usepackage{natbib}
\usepackage{soul} % in order to highlight
\usepackage{color}
\usepackage{longtable}
\usepackage{ textcomp }
\usepackage{verbatim}
\usepackage{fancyhdr} % Required for custom headers
\usepackage{lastpage} % Required to determine the last page for the footer
\usepackage{extramarks} % Required for headers and footers
\usepackage{graphicx} % Required to insert images
\usepackage{lipsum} % Used for inserting dummy 'Lorem ipsum' text into the template
\usepackage{ amssymb }
\usepackage{slashed}
\usepackage{tikz}
\usepackage{comment}
\usepackage[caption=false,listofformat=parens, subrefformat=parens]{subfig}
\usepackage{ marvosym }
\usepackage{array}
\usepackage{amsthm} % Theorem Formatting

\newcommand{\abs}[1]{\left| #1 \right|} % for absolute value
\let\baraccent=\= % rename builtin command \= to \baraccent
\renewcommand{\=}[1]{\stackrel{#1}{=}} % for putting numbers above =

\theoremstyle{definition}

\theoremstyle{remark}

\newcolumntype{C}[1]{>{\centering\let\newline\\\arraybackslash\hspace{0pt}}m{#1}}

\newcommand{\la}{\left <}
\newcommand{\ra}{\right >}

\newcommand{\nd}{n}
\begin{document}
\title{Mobility and quantum mobility of modern GaAs/AlGaAs heterostructures}
\author{M. Sammon} 
\email[Corresponding author: ]{sammo017@umn.edu} 
\affiliation{School of Physics and Astronomy, University of Minnesota, Minneapolis, MN 55455, USA}
\author{M. A. Zudov}
\affiliation{School of Physics and Astronomy, University of Minnesota, Minneapolis, MN 55455, USA}
\author{B. I. Shklovskii} 
\affiliation{School of Physics and Astronomy, University of Minnesota, Minneapolis, MN 55455, USA}

\received{\today}

\begin{abstract}
In modern GaAs/Al$_x$Ga$_{1-x}$As heterostructures with record high mobilities, a two-dimensional electron gas (2DEG) in a quantum well is provided by two remote donor $\delta$-layers placed on both sides of the well. 
Each $\delta$-layer is located within a narrow GaAs well, flanked by narrow AlAs layers which capture excess electrons from donors. 
We show that each excess electron is localized in a compact dipole atom with the nearest donor.
Nevertheless, excess electrons screen both the remote donors and background impurities. 
When the fraction of remote donors filled by excess electrons $f$ is small, the remote donor limited quantum mobility grows as $f^{3}$ and becomes larger than the background impurity limited one at a characteristic value $f_c$. 
We also calculate both the mobility and the quantum mobility limited by the screened background impurities with concentrations $N_1$ in Al$_x$Ga$_{1-x}$As and $N_2$ in GaAs, which allows one to estimate $N_1$ and $N_2$ from the measured mobilities. 
%Although at $f > f_c$ background impurities limit both mobilities, the quality of the quantum Hall effect at filling factor $\nu = 5/2$ is still determined by remote donors and thus is expected to improve with increasing $f$. 
%This improvement, however, can be hindered by spreading of donors in the GaAs doping well and roughness of the AlAs/GaAs interfaces.
Taken together, our findings should help to identify avenues for further improvement of modern heterostructures.
\end{abstract}

\maketitle

\maketitle

\section{Introduction} \label{sec:intro}

Modern GaAs/Al$_x$Ga$_{1-x}$As heterostructures hosting an ultra-high mobility two-dimensional electron gas (2DEG) are the result of glorious advances in molecular beam epitaxy.\cite{stormer:1979,pfeiffer:1989,umansky:1997,pfeiffer:2003,umansky:2009,UmanskyReview,ManfraReview,reichl:2014,gardner:2016} 
A more than $3000$ times increase of the electron mobility over the last several decades lead to numerous important discoveries, including odd-\cite{tsui:1982b} and even-\cite{willett:1987} denominator fractional quantum Hall effects, stripe and bubble phases,\cite{koulakov:1996,lilly:1999a,du:1999} exciton condensate in electron bilayers,\cite{eisenstein:2004} microwave-induced resistance oscillations and zero-resistance states,\cite{zudov:2001a,ye:2001,mani:2002,zudov:2003} etc. 
While there is growing experimental evidence that high mobility alone is not a good predictor of how a particular phenomenon manifests itself,\cite{umansky:2009,nuebler:2010,samkharadze:2011,pan:2011,gamez:2013,deng:2014,reichl:2014,ManfraReview,shi:2016b,qian:2017a} there clearly exists a strong interest in further improvement of GaAs/Al$_x$Ga$_{1-x}$As heterostructures.\cite{Sarma,DasSarma2014,DasSarma2015,ManfraReview,UmanskyReview,reichl:2014,gardner:2016}
It is therefore important to understand dominant sources of disorder and elucidate the ways to minimize them.

\begin{figure}[b]
	\begin{subfloat}
		{
			\includegraphics[width=\linewidth]{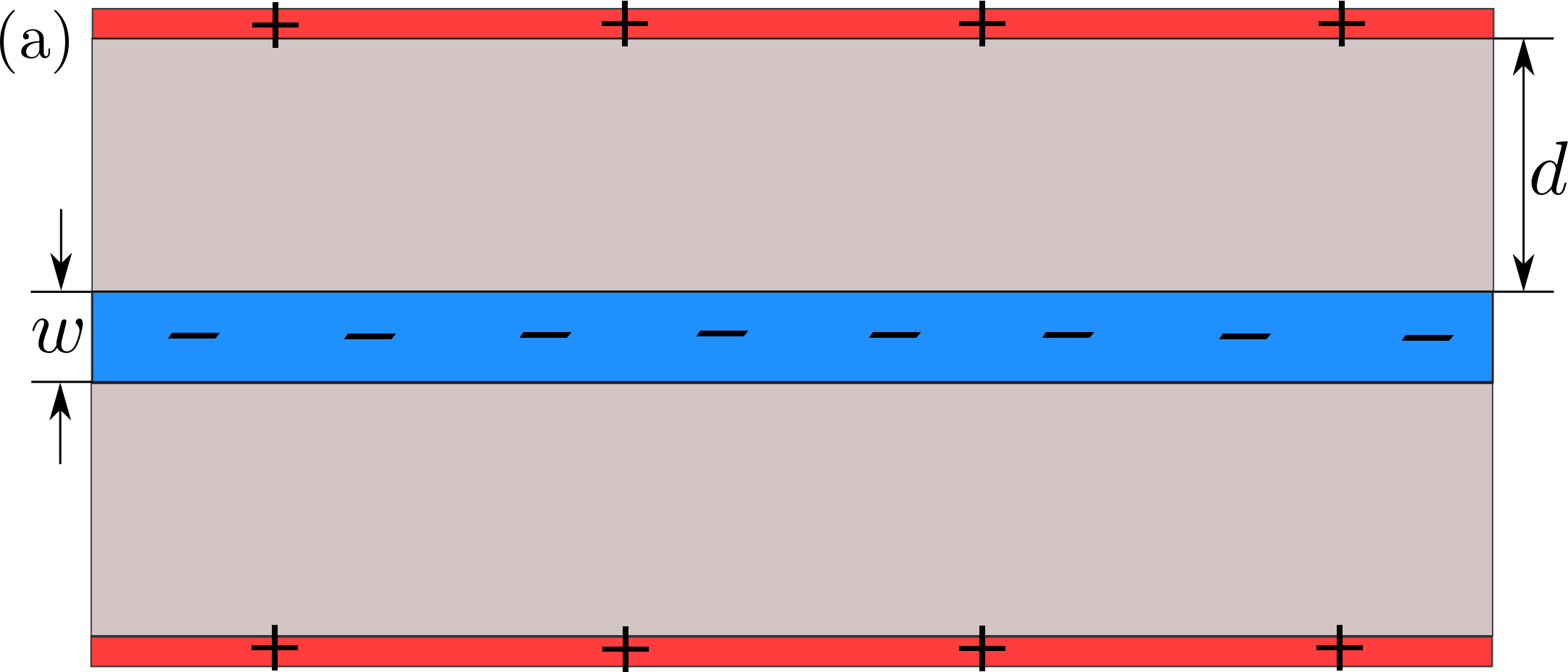}
			\label{fig:device}
		}
	\end{subfloat}
	\begin{subfloat}
		{
			\includegraphics[width=\linewidth]{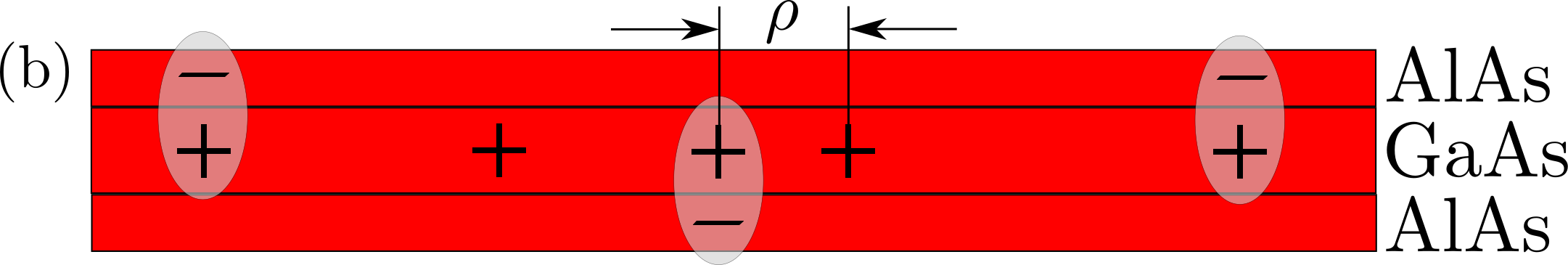}
			\label{fig:dopingwell}
		}
	\end{subfloat}
	\caption{(Color online) (a) A schematic view of a modern GaAs/Al$_x$Ga$_{1-x}$As heterostructure. 
		The 2DEG (shown in blue) resides in a GaAs well of thickness $w$ and is provided by two doping layers (shown in red) separated by Al$_x$Ga$_{1-x}$As barriers of thickness $d$ (shown in gray).
		Here, $-$ and $+$ represent negative and positive charges in the 2DEG and the doping layers, respectively.
		(b) An enlarged view of a small section of the doping layer at a filling fraction $f \simeq 0.6$. 
		Excess electrons ($-$) in AlAs form compact dipoles (ellipses) with the nearest donors ($+$) in GaAs. 
		Empty donors (also shown by $+$) alternate with compact dipoles due to Coulomb repulsion between the excess electrons. 
		Only empty donors are shown in Fig.\,1(a). 
		An example of a pair of anomalously close donors separated by distance $\rho$ is shown in the middle (see Sec.\,\ref{sec:remote}).}
\end{figure}

A typical modern GaAs/Al$_x$Ga$_{1-x}$As heterostructure, schematically shown in Fig.\,\subref{fig:device}, consists of a GaAs quantum well of width $w = 30$ nm surrounded by Al$_x$Ga$_{1-x}$As barriers. 
A 2DEG with a concentration $n_e \simeq 3 \times10^{11}$ cm$^{-2}$ fixed by the electrostatics of the device is provided by two remote doping layers symmetrically positioned at setback distances of $d\simeq 80$ nm. 
It has a low-temperature mobility $\mu \simeq 3 \times 10^7$ cm$^2$V$^{-1}$s$^{-1}$ and a quantum mobility\cite{note:0} $\mu_q \equiv e\tau_q/ m^\star \sim 1 \times 10^6$ cm$^2$V$^{-1}$s$^{-1}$,\cite{umansky:2009,shi:2016a,shi:2017a,qian:2017b,note:muqexp} where $\tau_q$ is the quantum lifetime and $m^\star \approx 0.067 m_e$ is the electron effective mass in GaAs.
Usually, $\mu$ and $\mu_q$ can be expressed as
\begin{align}
\mu^{-1}=\mu_{R}^{-1}+\mu_{B}^{-1}\,,\\
\mu_q^{-1}=\mu_{q,R}^{-1}+\mu_{q,B}^{-1}\,,
\end{align}
where the first and second terms account for scattering from remote ionized impurities (RI) and charged background impurities (BI), respectively. 
%In modern heterostructures these sources of scattering are minimized as follows. 
%First, . 
%Second, the remote doping layers have a sophisticated design which substantially reduces the electron scattering by ionized donors. 
In modern heterostructures, the BI concentration is extremely small ($\lesssim 10^{14}$ cm$^{-3}$) and the doping layers have a sophisticated design which substantially reduces RI scattering. 
As shown in Fig.\,\subref{fig:dopingwell}, each doping layer consists of a narrow (typical width of 3 nm) GaAs quantum well, which is doped in the middle by a $\delta$-layer of Si donors with a typical concentration $\nd\sim  10^{12}$ cm$^{-2}$ and surrounded by AlAs layers with a typical width of 2 nm. 
The doping layer shown in Fig.\,\subref{fig:dopingwell} is a special case of a short-period GaAs/AlAs superlattice (SPSL), suggested by Baba\cite{baba:1983} and later implemented by Friedland.\cite{Friedland}
Following Refs.\,[\onlinecite{UmanskyReview}], [\onlinecite{ManfraReview}], and [\onlinecite{Friedland}] we use this abbreviation for the doping scheme shown in Fig.\,\subref{fig:dopingwell}. 
To our knowledge, a structure shown in Fig.\,1 was first realized in the early 2000s,\cite{cooper:2001,eisenstein:2002} although single heterointerfaces with one such doping layer appeared much earlier.\cite{du:1993,pfeiffer:2018}

The SPSL-doping scheme augments the advantage of weak RI scattering of the 2DEG electrons. 
In AlAs/GaAs heterostructures with thick layers, the X-minima in AlAs are higher in energy than the $\Gamma$-minimum in GaAs. However, for thin layers size quantization plays an important role. In our SPSL-doping layers, the much lighter effective mass in GaAs produces a much larger size quantization energy than in AlAs, raising the $\Gamma$-minimum in GaAs above the X-minima in AlAs.\cite{Ihm, Reichl2018} 
Thus, all donated electrons which are not transferred to the 2DEG (excess electrons) are stored in the AlAs side wells. 
Each excess electron pairs with a donor in a compact dipole atom, so that their low-temperature parallel-to-2DEG conductance is negligible. 
Furthermore, excess electrons choose donors which minimize their energy; this leads to significant correlations in the positions of charged donors\cite{Efros1,Dohler} and thus to a dramatic reduction of RI scattering. 
In what follows, we call this redistribution of electrons excess electron screening (EES) and describe it by the screening radius $r_s$.
EES is different from the conventional screening by the 2DEG which is described in the paper by the Thomas Fermi (TF) screening radius $q_{TF}^{-1}$ and exists on top of the EES. 
As we show, EES also reduces the BI potential. 

In the first part of this paper we study scattering by the charged donors in the SPSL-doping layers. 
%We begin with the calculation of the binding energy of a compact dipole atom and show that the electron localization length in the middle of the AlAs layers is 2.7 nm, i.e., nearly four times smaller than the average distance 10 nm between the donors. 
We begin with the calculation of the binding energy of a dipole atom and show that the electron localization length in the middle of the AlAs layers is 2.7 nm, several times smaller than the average distance between the donors.
This means that the excess electrons are indeed localized and at low temperature their parallel-to-2DEG conductance is negligible. Localization of the excess electrons also allows us to treat the doping layer as a lightly doped 2D semiconductor and to use the classical theory of its ground state\cite{Bello,Efros1} to study the EES. 
 We then calculate $\mu_R$ and $\mu_{q,R}$ limited by a single doping layer containing donors with concentration $n$ and excess electrons with concentration $fn$, where $f$ is what we call the donor filling fraction. In the device shown in Fig. 1(a), neutrality requires that $f=1- n_e/2n$ and $f$ can be varied by changing $n$. In addition, some electrons can be lost to the device surface (not shown) and $f$ can be different even when $n$ is the same. Thus, for our analysis we treat $f$ as an independent variable. 

We find that $\mu_R$ and $\mu_{q,R}$ grow very fast with $f$ and exceed $\mu_B$ and $\mu_{q,B}$ in the range $0.15<f<0.39$. In this range we find 
\begin{align}
\mu_R \simeq 24 f^3 \frac{e}{\hbar}k_F^3d_w^5\,,
\label{eq:mobility_final}\\ 
\mu_{q,R} \simeq 24f^3\frac{e}{\hbar}k_Fd_w^3\,, \label{eq:quantummobility_final}
\end{align}
where $k_F = (2\pi n_e)^{1/2}$ is the Fermi wavenumber of the 2DEG and $d_w\equiv d+w/2$ is the distance between the center of the quantum well and the doping layers.

Eqs.\,(\ref{eq:mobility_final}) and (\ref{eq:quantummobility_final}) are valid only if they predict mobilities larger than the standard values in the presence of $\nd$ donors and no excess electrons ($f=0$),\cite{price,Gold,dmitriev:2012}
\begin{align}
\mu_R=\frac{8e}{\pi\hbar}\frac{(k_Fd_w)^3}{\nd}\label{eq:mobilityAndo_approx}\,,\\
\mu_{q,R}=\frac{2e}{\pi\hbar}\frac{k_Fd_w} {\nd}\label{eq:quantummobilityAndo_approx}\,.
\end{align}

%For $n_e = 3 \times10^{11}$ cm$^{-2}$, $\nd = 1 \times 10^{12}$ cm$^{-2}$, and $d_w = 95$ nm, Eqs.\,(\ref{eq:mobilityAndo_approx}) and (\ref{eq:quantummobilityAndo_approx}) result in $\mu_R \simeq 8 \times 10^6$ cm$^2$V$^{-1}$s$^{-1}$ and $\mu_{q,R} \simeq 1 \times 10^4$ cm$^2$V$^{-1}$s$^{-1}$.
%These values are significantly lower than total experimental mobilities in modern heterostructures. 

Because the doping layers in real samples likely have different $f$, for the estimates below we use Eqs.\,(\ref{eq:mobility_final}) and (\ref{eq:quantummobility_final}) for a single layer with the smallest $f$. Unless otherwise specified, here and below we use the reference sample parameters\cite{note:pars} $n_e = 3 \times10^{11}$ cm$^{-2}$, $d_w = 95$ nm, $\mu = 3 \times 10^7$ cm$^2$V$^{-1}$s$^{-1}$, and $\mu_q = 1 \times 10^6$ cm$^2$V$^{-1}$s$^{-1}$.
From Eqs.\,(\ref{eq:mobility_final}) and (\ref{eq:quantummobility_final}) we find that $\mu_R=\mu_B$ at $f \approx 0.20$, while $\mu_{q,R}=\mu_{q,B}$ at $f = f_c \approx 0.36 $.\cite{note:1}
Since $f$ in modern heterostructures often exceeds this value (see, e.g., Ref.\,[\onlinecite{umansky:2009}]), this result suggests that not only $\mu$, but also $\mu_q$, can be governed by BI scattering.

\iffalse
{\bf Thus, the optimization of $\mu_q$ requires that every doping layer keeps more than 31\% of its electrons. For the idealized symmetric neutral structure shown in Fig.\,\subref{fig:device} this is not a problem because $f=1- n_e/2n=0.85$ and remote donors should play no role for either $\mu$ or $\mu_q$. However, in many samples surface states ``steal" electrons from the top SPSL layer and reduce $f$. Three ways to prevent this are being used.\cite{UmanskyReview,ManfraReview} First, the compensation of the surface by an additional close to surface SPSL. Second, the surface is moved further from the doping layer, as the concentration of electrons captured by the surface is inversely proportional to the distance between the doping layer and the surface. A third way is to increase the donor concentration $\nd$ in the doping layer closest to the surface. Still when the surface is taken care of the typical structures are asymmetric so that one doping layer has a smaller $f$ and determines the RI-limited mobility. Therefore, for the estimates above we used Eqs.\,(\ref{eq:mobility_final}) and (\ref{eq:quantummobility_final}) for a single layer with the smallest $f$.}
\fi

%In the second part of the paper we assume that the condition $f > f_c$ is fulfilled and calculate mobilities limited by background impurities. 
In the second part of the paper, we calculate $\mu_B$ and $\mu_{q,B}$ taking into account EES.
We show that at $f\geq f_c$ EES eliminates scattering by all impurities located at distances larger than $0.5d_w$ from the midplane of the 2DEG's quantum well.
We assume BI concentrations $N_1$ and $N_2$ in the Al$_x$Ga$_{1-x}$As barriers and GaAs quantum well, respectively, and arrive at the linear equations 
\begin{align}\label{eq:mobility_linearnumbers}
\mu_B^{-1}=A_1N_1+A_2N_2\,,\\
\mu_{q,B}^{-1}=B_1N_1+B_2N_2\,,\label{eq:quantummobility_linearnumbers}
\end{align}
with coefficients $A_1 \ll A_2$ and $B_1 \sim B_2$ which we calculate.
As a result, Eqs.\,(\ref{eq:mobility_linearnumbers}) and (\ref{eq:quantummobility_linearnumbers}) allow one to estimate $N_1$ and $N_2$ from measured $\mu_B$ and $\mu_{q,B}$. 
We find\cite{note:pars} $N_1 \simeq 2\times 10^{14}$ cm$^{-3}$, $N_2\simeq2\times10^{13}$ cm$^{-3}$.
These estimates suggest that $\mu_{q,B}$ is dominated by BI in the Al$_x$Ga$_{1-x}$As barriers, due to their larger concentration, and therefore should benefit from the purification of the Al source.\cite{reichl:2014,LorenAl}

The plan of this paper is as follows. 
In Sec.\,\ref{sec:variation} we study the quantum mechanics of an isolated compact dipole atom in the doping layer. 
We compute the binding energy of an electron in a compact dipole atom and show that its localization length in the plane of the layer is small enough to proceed classically. 
In Sec.\,\ref{sec:remote} we study the screening of fluctuations of the donor concentration $\nd(\rho)$ by EES and compute $\mu_R$ and $\mu_{q,R}$ [Eqs.\,(\ref{eq:mobility_final}) and (\ref{eq:quantummobility_final})]. 
In Sec.\,\ref{sec:} we compute $\mu_B$ and $\mu_{q,B}$ [Eqs.\,(\ref{eq:mobility_linearnumbers}) and (\ref{eq:quantummobility_linearnumbers})], taking into account EES of the BI and derive simple analytical formulas for $A_1$, $A_2$, $B_1$, and $B_2$.
In Sec.\,\ref{sec:disorder} we examine the possible suppression of EES by spreading of the $\delta$-layer of Si donors in the GaAs well and by roughness of the AlAs/GaAs and AlAs/Al$_x$Ga$_{1-x}$As interfaces. 
In Sec.\,\ref{sec:5/2QHE} we comment on the possible relation between the RI potential and the measured gap of the fractional quantum Hall effect at filling factor 5/2. 
We conclude in Sec.\,\ref{sec:summary} with a summary of our results and possible avenues for further improvement of GaAs/Al$_x$Ga$_{1-x}$As heterostructures.

\section{Localization of electrons in the doping layers}\label{sec:variation}

A remarkable feature of the SPSL-doping scheme is that the excess electrons in the AlAs layers are able to reduce the random potential of donors in the GaAs layer but their parallel-to-2DEG conductance is negligible. 
As stated above, the main reasons for this are the proximity of the electrons to the donors and the large effective electron mass in AlAs. 
In this section we justify this claim, showing that excess electrons, while residing in AlAs, are strongly bound to donors in GaAs. 
\begin{figure}[h!]
	\includegraphics[width=0.8\linewidth]{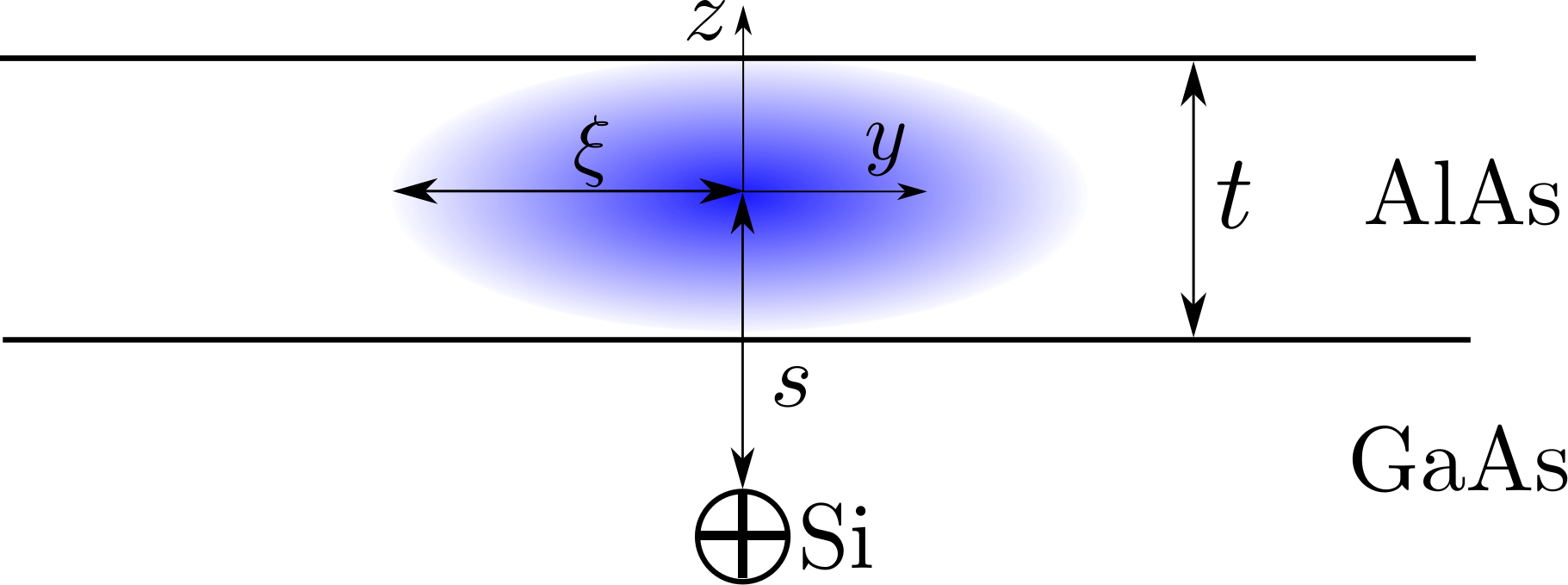}
	\caption{(Color online) Schematic image of the electron wave function cloud (blue) in the AlAs layer of thickness $t$.
This cloud is bound to a Si donor ($+$) in the GaAs layer a distance $s$ away from the midpoint of the AlAs layer. 
$\xi$ is the electron localization length in the $x-y$ plane.
% defined in Eq.\,(\ref{eq:localization_length}).
}\label{fig:boundstate}
\end{figure}

An illustration of an electron bound state is shown in Fig.\,\ref{fig:boundstate}. 
Each excess electron resides in the middle of the AlAs layer of thickness $t$ and is bound to a donor in GaAs at a distance $s$ away with the localization length $\xi$ in the $x-y$ plane. 
%The goal of this section is to estimate the product $\nd\xi^2$ which measures the overlap between electron wave functions bound to neighboring donors and show that $\nd\xi^2$ is so small that strong localization condition is satisfied.
The $z$-axis is perpendicular to the AlAs/GaAs interface and the origin is centered above the donor at the midpoint of the AlAs layer.
%, so that in AlAs $z$ varies from $-t/2$ to $t/2$. 
The AlAs/GaAs and AlAs/Al$_x$Ga$_{1-x}$As interfaces are treated as infinite barriers so that the electrons are completely confined to the AlAs layer. 
This means that there are two competing energy scales: the separation $\Delta$ between the first and the second subbands of the AlAs layer and the Coulomb binding energy $E_b$. 
Below we show that $E_b \ll \Delta$ for reference sample parameters $t=2$ nm and $s=2.5$ nm. 
This allows us to think that electrons are bound in the plane at $z=0$ by an effective 2D potential
%The effective potential can be written as
 \begin{equation}
 V(\rho,s)=-\frac{2}{t}\int\limits_{-t/2}^{t/2}dz\cos^2\left(\frac{\pi z}{t}\right)\frac{e^2}{\bar{\kappa}\sqrt{\rho^2+(s+z)^2}}\,,
 \end{equation}
 obtained by averaging the Coulomb attraction of the donor over the ground state wave function $\phi(z)=(2/t)^{1/2}\cos(\pi z/t)$. Here $\rho = \sqrt{x^2+y^2}$ 
%is the radius in the $x-y$ plane 
and $\bar{\kappa}$ is the effective dielectric constant. Because the dielectric constants of GaAs ($\kappa\simeq13$) and AlAs ($\kappa_{\mbox{\scriptsize{A}}}\simeq10$) are relatively close, we use $\bar{\kappa}= (\kappa+\kappa_{\mbox{\scriptsize{A}}})/2\simeq11.5$. (Here and below we do not discriminate between the dielectric constants of GaAs and Al$_{x}$Ga$_{1-x}$As for the relevant $x\simeq0.24$.) 
The corresponding Schr\"{o}dinger equation is then given by 
\begin{equation}
-\frac{\hbar^2}{2m_{xy}^\star }\nabla^2\psi(\rho)+V(\rho,s)\psi(\rho)=-E_b\psi(\rho)\,,
\end{equation}
where $m_{xy}^\star $ is the electron's effective mass in the $x-y$ plane. 
To find $E_b$ we use a variational approach with the trial wave function
\begin{equation}\label{eq:wavefunction}
\psi(\rho)=\exp\left(-\frac{\sqrt{\rho^2+s^2}}{b}\right)\,,
\end{equation}
where $b$ is the variational parameter which minimizes $E_b$. 

%This function matches the approximate solutions of the 2D Schr\"{o}dinger equation at $\rho\ll s$ and $\rho\gg s$ when $t=0$. Indeed, at $\rho\ll s$ we expand the potential of the electron and find $V(\rho,s)\approx -e^2/s+e^2\rho^2/(2s^3)$. 
%This creates a quadratic radial potential, and thus at small $\rho$ the wave function is Gaussian in the radial distance $\rho$, $\psi\propto \exp[-\rho^2/(2bs^3)]$. Conversely, at $\rho\gg s$ Eq.\,(\ref{eq:wavefunction}) gives $\psi\propto\exp(-\rho/b)$ as in the ground state wavefunction of a 2D hydrogen atom. Thus, our trial wave function is expected to give reasonable estimates of the binding energy of the electrons. 

\begin{figure}
	\includegraphics[width=\linewidth]{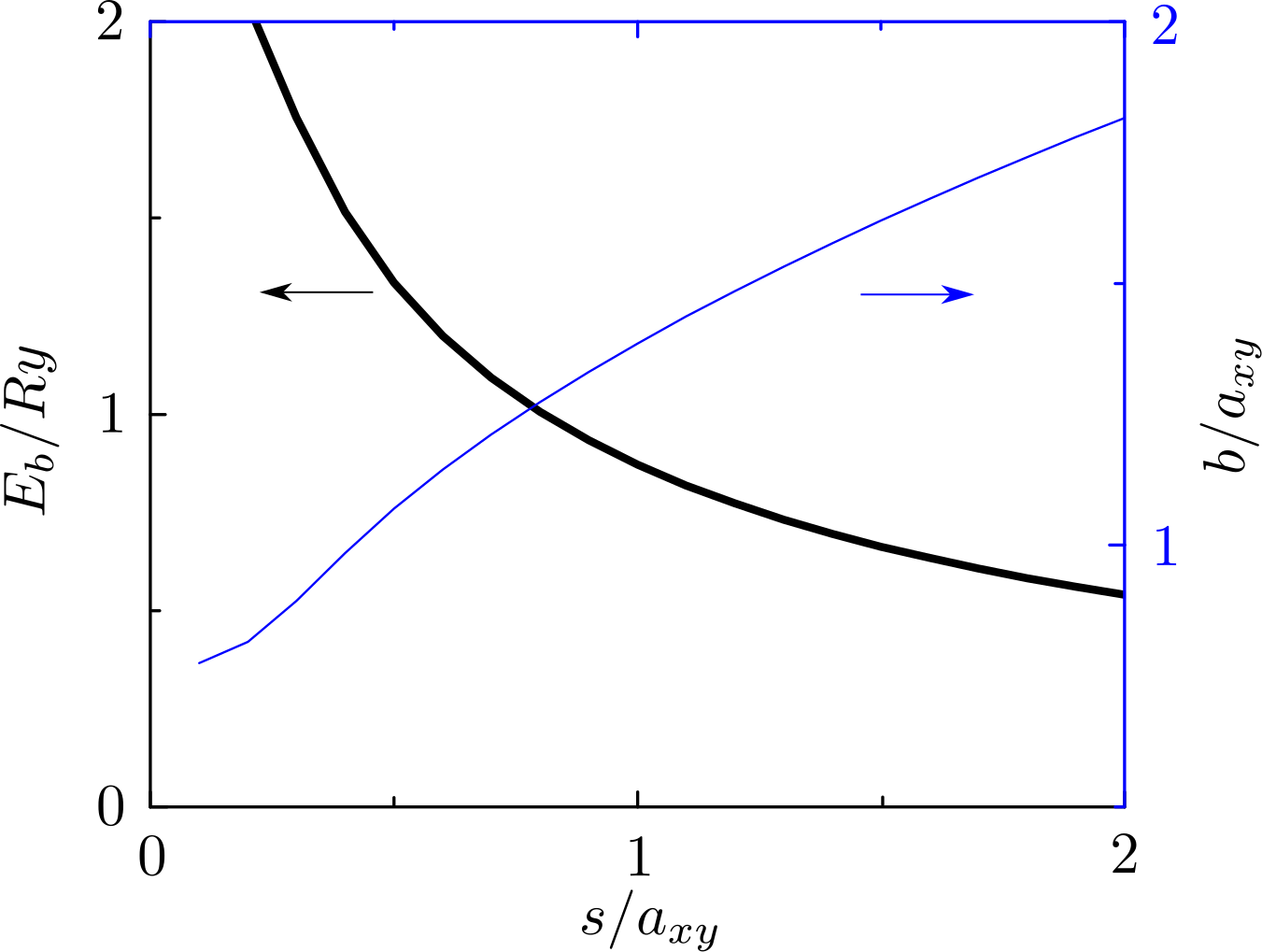}
		\caption{(Color online) The binding energy $E_b$ in $Ry$ (thick line) and the variational parameter $b$ in units of the in-plane effective Bohr radius $a_{xy}$ (thin line) as a function of the distance $s$ to the binding donor in units of $a_{xy}$ as obtained from the variational calculation for $t=2$ nm. }
	\label{fig:variation}
\end{figure}

The results of the variational calculation $E_b/Ry$ and $b/a_{xy}$ as a function of $s/a_{xy}$ are given in Fig.\,\ref{fig:variation}, where $a_{xy}=\bar{\kappa}\hbar^2/(m_{xy}^\star e^2)$ is the in-plane effective Bohr radius and $Ry=\hbar^2/(2m_{xy}^\star a_{xy}^2)$.
Using $m_{xy}^\star = 0.22 m_e$ for AlAs, we find $a_{xy} \simeq 2.6$ nm and $Ry \simeq 23$ meV near the $X$-point minima.\cite{Vurgaftman} 
With $s=2.5$ nm, we then estimate the electron binding energy $E_b \approx 21$ meV. 

Above we have assumed that $E_b \ll \Delta$, allowing us to average the potential over the fast motion along $z$-direction and treat an electron as two-dimensional. 
To justify this assumption we estimate the inter-subband separation $\Delta=3\hbar^2\pi^2/(2m_z^\star t^2)$. 
The electronic spectrum near the $X$-point minima in AlAs is anisotropic and $ m^\star_z=0.95 m_e$.\cite{Vurgaftman} 
We find that indeed $\Delta \simeq 0.26$ eV $\gg E_b$.

The localization length $\xi$ of the electron in an isolated dipole atom in the $x-y$ plane is given by 
\begin{equation}\label{eq:localization_length}
\xi=\frac{\hbar}{\sqrt{2m_{xy}^\star E_b}}\,,
\end{equation}
which yields $\xi \simeq 2.7$ nm. 
For $\nd=1\times 10^{12}$ cm$^{-2}$, we find $\nd \xi^2 \simeq 0.07$ which should be compared to the critical value of $(\nd\xi^2)_c$ below which electrons are localized and transport is activated.

We can estimate $(\nd\xi^2)_c$ using the data for a Si MOSFET doped by sodium at the SiO$_2$ side of the interface.\cite{fowler:1980} 
These sodium atoms donate electrons which reside on the silicon side of the interface. 
Such MOSFET is therefore similar to the SPSL-doping layer in which a sodium ion in SiO$_2$ assumes the role of Si in GaAs.
The activation energy $E_1$ of the electron conductivity along the interface $E_1(\nd)$ as a function of the surface concentration of sodium $\nd$ was investigated in Ref.\,[\onlinecite{fowler:1980}]. 
At small $\nd$, $E_1 \approx 24$ meV, in agreement with theoretical predictions for the binding energy of an isolated donor. 
With increasing $\nd$, $E_1 (\nd) $ decreases and extrapolation to large $\nd$ shows that it vanishes at $\nd \approx 1.7\pm0.5\times 10^{12}$ cm$^{-2}$. 
Using Eq.\,(\ref{eq:localization_length}) with the binding energy of an isolated donor (24 meV) and the in-plane effective electron mass ($0.19$ $ m_e$) we find the localization length of the electron bound to an isolated sodium ion $\xi \approx 2.9$ nm and conclude that in Si MOSFET localization sets in at $(\nd\xi^2)_c \approx 0.14\pm0.04$.
Since our estimate of $\nd \xi^2 \approx 0.07$ for SPSL-doping layer is smaller than this value, excess electrons should be localized and their hopping conductivity at low $T$ should be much smaller than $e^2/h$ (and activated).\cite{note:2}
Measurements of the conductivity of SPSL-doping layers have shown that it is indeed activated.\cite{DoroEng,dorozhkin:2018}

\section{Scattering by remote donors}\label{sec:remote}

Since excess electrons in the AlAs layers and donors in the GaAs layer form compact dipole atoms, scattering from these dipoles can be ignored. 
However, localized electrons can still choose among host donors, minimizing the total energy of the system. 
As a result, the ionized donors are screened by the $f\nd$ electrons, so that the correlator of the random potential energy $\la U(\boldsymbol{\rho})U(0)\ra$ is reduced ($\boldsymbol{\rho} =(x,y)$ is a vector in the $x-y$ plane).
The Fourier image of the potential correlation function and the Fourier image of the correlator $D(\rho)\la \nd(0)[1-f(0)]\nd(\boldsymbol{\rho})[1-f(\boldsymbol{\rho})]\ra$ of ionized donor concentration fluctuations can be related as
\begin{equation}\label{eq:potential_correlator} 
\la|U(q)|^2\ra=\left(\frac{2\pi e^2}{\kappa q}\right)^2D(q),
\end{equation}
and so the screening of the potential can be understood as originating from the correlations of the ionized donors. If donors of concentration $n$ are charged and uncorrelated, $D(q)=n$ in Eq. (\ref{eq:potential_correlator}). At small $f$ when the concentration of ionized donors is still approximately $n$, screening by the excess electrons reduces $\la|U(q)|^2\ra$ (and thus $D(q)$) by the factor $(1+(qr_s)^{-1})^{2}$. Accounting for the additional factor $(1-e^{-2qd_w})^{2}$ from the images of the donors in the 2DEG, $D(q)$ can be written as 
\begin{equation}\label{eq:generalchargecorrelator}
D(q) \simeq \frac {(qr_s)^{2}\nd}{(1-e^{-2qd_w})^{2}}\,,
\end{equation}
where we have used the condition $qr_s \ll1$, valid for the important wave numbers $q\sim d_w^{-1}$ and not too small $f$ (see below).

Since $d_w\gg w$ we can treat the 2DEG as if it were confined to an infinitely thin plane located at the center of the quantum well. 
The contributions of RI scattering to the mobility and quantum mobility can then be calculated using Born approximation as
\begin{gather}\label{eq:mobility}
\mu_R^{-1}=\frac{ 2\pi\hbar}{e a_B^2}\int\limits_0^{2\pi}\frac{d\theta (1-\cos\theta)e^{-2qd_w}}{(q+q_{TF})^2}D(q)\,,\\ \label{eq:quantummobility}
\mu_{q,R}^{-1}=\frac{2\pi \hbar}{ea_B^2}\int\limits_0^{2\pi}\frac{d\theta e^{-2qd_w}}{(q+q_{TF})^2}D(q)\,,
\end{gather}
where $q=2k_F\abs{\sin(\theta/2)}$ is the transferred momentum, $\theta$ is the angle between the initial electron wave vector $\textbf{k}$ and the final wave vector $\textbf{k}+\textbf{q}$, $q_{TF}=2a_B^{-1}$ is the inverse Thomas-Fermi screening radius of the 2DEG, and $a_B=\kappa\hbar^2/m^\star e^2 \simeq 10$ nm is the effective Bohr radius in GaAs. 

The main contribution to the integrals in Eqs.\,(\ref{eq:mobility}) and (\ref{eq:quantummobility}) comes from $q \lesssim (2d_w)^{-1}$. 
For such $q$, $q+q_{TF}\simeq q_{TF}$ (since $a_B \ll d_w$). 
Changing the integration variable to $q$, and extending the upper bound of integration to $\infty$ (since $k_Fd_w\gg1$), we find
\begin{align}
\mu_R^{-1}=\frac{\pi\hbar}{2ek_F^3} \int\limits_0^{\infty}q^2D(q)e^{-2qd_w}dq\label{eq:mobility_approx}\,,\\
\mu_{q,R}^{-1}=\frac{\pi\hbar}{ek_F} \int\limits_0^{\infty}D(q)e^{-2qd_w}dq\label{eq:quantummobility_approx}\,.
\end{align} 

In Eqs.\,(\ref{eq:mobility}), (\ref{eq:quantummobility}), (\ref{eq:mobility_approx}), and (\ref{eq:quantummobility_approx}) the random fluctuations of the RI potential are screened twice: once by EES and once by 2DEG screening. 
In the absence of EES ($f=0$), a single layer of donors with concentration $\nd$ is characterized by  $D(q)=\nd$ and one arrives at Eqs.\,(\ref{eq:mobilityAndo_approx}) and (\ref{eq:quantummobilityAndo_approx}) with the well-known ratio $\mu_R/\mu_{q,R} = (2k_{F} d_w)^2$.

We now return to Eq.\,(\ref{eq:generalchargecorrelator}) and calculate $r_s$ for $f\ll1$.
Since only a small concentration $f n$ of excess electrons remain in the AlAs barriers, they only occupy the lowest energy states. 
Such states are provided by rare pairs of anomalously close donors, separated by distance $\rho$ (see such a pair in Fig.\,\subref{fig:dopingwell}). 
An electron forms a dipole with one donor while the other donor remains ionized and its attractive potential lowers the electron energy by $e^2/\bar{\kappa} \rho$ when $\rho\gg s$, $\xi$.  
At small $f$, one can easily calculate the chemical potential $E_F(f,\nd)$ which separates the energy levels of the occupied and empty dipole atoms, and is measured from the energy of an isolated dipole atom.\cite{Bello}
The probability to find a second donor in a disk of radius $\rho_F\equiv e^2/\bar{\kappa} \abs{E_F}$ centered around the first donor is $\pi \nd\rho_F^2$. 
The average concentration of such donor pairs, i.e., the concentration of electrons, is then $\pi \nd^{2} \rho_F^2 / 2 = f \nd$, where factor 1/2 eliminates double counting. 
Recalling that $E_F$ is negative, one then finds\cite{Bello}
\begin{equation}\label{eq:EF_1-K}
E_F=-\left(\frac{\pi}{2}\right)^{1/2}\frac{e^2 \nd^{1/2}}{\bar{\kappa}}\frac{1}{f^{1/2}}\,,
\end{equation}
and
\begin{equation}\label{eq:screeningradius}
r_s =\frac{\bar{\kappa}}{2\pi e^2} \frac{1}{\nd}\frac{dE_F}{df} = \frac{1}{4(2\pi \nd)^{1/2}}\frac{1}{f^{3/2}}\simeq\frac{0.1}{\nd^{1/2}f^{3/2}}\,.
\end{equation} 
Unfortunately, Eqs.\,(\ref{eq:EF_1-K}) and (\ref{eq:screeningradius}) are only valid for very small $f\ll0.15$.\cite{Bello,note:9} 
As mentioned in the introduction, the mobilities likely cross over from being limited by RI scattering to being determined by BI scattering in the range $0.15<f<0.39$. 
In this range, we use the results of numerical modeling of the ground state of the $f\nd$ excess electrons on $\nd$ random donors and a neutralizing background.\cite{Bello,Efros1} 
For $0.15<f<0.39$ we find a simple fit
\begin{equation}\label{eq:screeningradiusnumerical}
r_s\simeq\frac{0.18}{\nd^{1/2}f^{3/2}}\,,
\end{equation}
with an accuracy of 20\%. Using this $r_s$ with Eq.\.(\ref{eq:generalchargecorrelator}) and combining with Eqs.\,(\ref{eq:mobility_approx}) and (\ref{eq:quantummobility_approx}), we arrive at Eqs.\,(\ref{eq:mobility_final}) and (\ref{eq:quantummobility_final}).\cite{note:10} 

\begin{comment}
Furthermore, for this range of $f$ we can no longer assume that the concentration of randomly positioned charged donors is $n$. Instead, we must reduce $n$ in Eq.\.(\ref{eq:generalchargecorrelator}) by the factor $(1-f)^3$. Indeed, only $(1-f)n$ donors remain charged and their spacial distribution is correlated because the compact pairs of charged donors with $\rho < \rho_F$ have been eliminated. For small $q\sim 1/d$ this results in the effective concentration of random charged donors further reduced by $1-\pi(1-f)n\rho_F^2 \simeq
(1-f)^2$.\cite{DasSarma2015} Using this reduced concentration in Eq.\.(\ref{eq:generalchargecorrelator}) with Eq.\.(\ref{eq:screeningradiusnumerical}), and combining with Eqs.\,(\ref{eq:mobility_approx}) and (\ref{eq:quantummobility_approx}), we arrive at Eqs.\,(\ref{eq:mobility_final}) and (\ref{eq:quantummobility_final}). 
\end{comment}

Our results are based on the assumption that the donors are randomly distributed in the plane of their $\delta$-layer. 
The distribution of donors at low temperatures is a snapshot of the distribution of donors at a temperature $T_{D} \sim 800 {\rm~K}\sim 6 e^{2}\nd^{1/2}/\bar{\kappa}$ below which the diffusion of donors stops. 
At this temperature dipole atoms are ionized and donors separated by a distance $\rho$ interact with the Coulomb repulsion energy $e^2/\bar{\kappa} \rho$. 
If $e^2/\bar{\kappa} \rho > T_{D}$, the probability to find such a pair of donors is reduced by the Boltzmann factor $\exp[- (e^2/\bar{\kappa} \rho)/T_{D}]$. 
For the important pairs, $e^2/\bar{\kappa} \rho_F \sim |E_F|$, and using Eq.\,(\ref{eq:EF_1-K}) we find that this effect is relevant only at $f \lesssim 0.05$, where EES plays little role. Therefore, Eqs.\,(\ref{eq:mobility_final}) and (\ref{eq:quantummobility_final}) are robust against this effect for experimentally relevant $f$.

We have also assumed that the system of excess electrons is close to its ground state at low temperatures. 
Although these electrons are localized, the rate of electron hops from a dipole atom to the nearest empty donor $\gamma \simeq \gamma_0 \exp[-2(\nd\xi^2)^{-1/2}]$ has a large prefactor $\gamma_0 \sim 10^{12}$ s$^{-1}$ related to the emission of phonons.
% with energy $ E_F \sim 200$ K. 
For $\nd = 1\times 10^{12}$ cm$^{-2}$ the exponential factor is $\sim 10^{-3}$ resulting in $\gamma \sim 10^{9}$ s$^{-1}$, much larger than the typical rate of cooling of the sample.

Our Eq.\,(\ref{eq:mobility_final}) for $\mu_R$ can be compared with Ref.\,[\onlinecite{Efros1}] which numerically studied the screening of the RI potential by excess donors in GaAs/Al$_x$Ga$_{1-x}$As heterostructures with a conventional $\delta$-doping in Al$_x$Ga$_{1-x}$As at $d \leq 50$ nm (only the equilibrium theory of Ref.\,[\onlinecite{Efros1}] is relevant here).\cite{fnoteDasSarma} 
In the important range of filling fractions $0.2 < f < 0.4$ if we use the parameters of Ref.\,[\onlinecite{Efros1}] our $\mu_R$ agrees with its Fig.\,4(a). 
Ref.\,[\onlinecite{Efros1}] did not study $\mu_q$ or BI scattering. 

As we saw above, in modern heterostructures both $\mu_R$ and $\mu_{q,R}$ are larger than $\mu_B$ and $\mu_{q,B}$ at $f > f_c = 0.36$.
However, the remote donors can become important for the quantum mobility if one succeeds to substantially reduce BI scattering. 
We thus would like to estimate $\mu_R$ and $\mu_{q,R}$ at $1-f \ll 1$, i.e., when almost all of the donors form neutral dipole atoms and only a small fraction of donors $1-f \ll 1$ are ionized. 
Ionized donors can be treated as holes which repel each other and tend to form a Wigner crystal.\cite{Bello,Dohler} 
If such a crystal were ideal, it would not scatter electrons. 
However, due to the discreteness of the random positions of donors, holes have to move from their ideal position to the nearest neighbor donor. 
Each such move effectively creates a dipole with the arm $\sim \nd^{-1/2}$ randomly oriented in the $x-y$ plane. 
The number of such dipoles in the relevant square of size $d_w$ is $(1-f) \nd d_w^2$ and because of their random orientation the amplitude of potential fluctuations created by them in the 2DEG can be estimated as $[(1-f) \nd d_w^2]^{1/2}(e \nd^{-1/2}/\kappa d_w^2) = (1-f)^{1/2}(e/\kappa d_w)$. 
%This is a very small potential which cannot be screened by localized excess electrons.\cite{Baranovskii} 
A more accurate estimate of the dipole scattering gives, 
\begin{align}\label{eq:mu_R:WC}
\mu_R\simeq7.7\frac{e}{\hbar}\frac{k_F^{3}d_w^{5}}{1-f}\,,\\
\mu_{q,R}\simeq6.5\frac{e}{\hbar}\frac{k_Fd_w^{3}}{1-f}\,.\label{eq:mu_qR:WC}
\end{align} 
Note, that our results for $1-f \ll 1$ disagree with Ref.\,[\onlinecite{Efros1}], which arrived at a much faster growth of $\mu_R$ near $f=1$.
This is because the authors only considered macroscopic fluctuations of the donor concentration and ignored fluctuations in the position of the nearest neighbors mentioned above.

\begin{comment}
At $f \gtrsim 0.5$, where the random potential is of the order of magnitude of a single donor potential $e/\kappa d_w$, the mobilities $\mu_R$ and $\mu_{q,R}$ crossover back to Eqs.\,(\ref{eq:mobility_final}) and (\ref{eq:quantummobility_final}).
\end{comment}

\section{Scattering by background impurities}\label{sec:}
In this section we consider scattering by background impurities in SPSL-doped heterostructures taking into account EES. 
We begin with $\mu_B$ and $\mu_{q,B}$ calculated in the Born approximation as
 \begin{align}\label{eq:mobilityBG}
 \mu_B^{-1}=\frac{ m^{\star 2}}{e\pi\hbar^3k_F^2}\int\limits_0^{2k_F}\la|U(q)|^2\ra\frac{q^2}{\sqrt{4k_F^2-q^2}}dq\,,\\ \label{eq:quantummobilityBG}
 \mu_{q,B}^{-1}=\frac{2m^{\star 2}}{e\pi\hbar^3}\int\limits_0^{2k_F}\la|U(q)|^2\ra\frac{1}{\sqrt{4k_F^2-q^2}}dq\,,
 \end{align}
 where $\la|U(q)|^2\ra$ is the BI scattering potential. 
 
 In this section as everywhere above we assume that the 2DEG occupies the first subband only. Then the square of the wave function is well concentrated in the range 
$-w/4< z< w/4$ near the midplane of the quantum well. 
If the BI are uniformly distributed with a concentration $N$, then in the absence of EES, the scattering potential can be written as
 \begin{equation}\label{eq:_unscreened}
 \la|U(q)|^2\ra=\frac{N}{q}\left(\frac{2\pi e^2}{\kappa (q+q_{TF})}\right)^2\,.
 \end{equation}
It is easy to see that with such a $\la|U(q)|^2\ra$ Eq.\,(\ref{eq:quantummobilityBG}) diverges logarithmically. This divergence results from scattering from an infinite number of distant impurities. (These impurities scatter at small angles so that they do not affect transport, and $\mu_B^{-1}$ remains finite.) In order to truncate this divergence, one must either consider a sample with finite thickness, or use multiple scattering theory that goes beyond the Born approximation.\cite{Macleod, Gold}
 
\begin{figure}
 	\includegraphics[width=\linewidth]{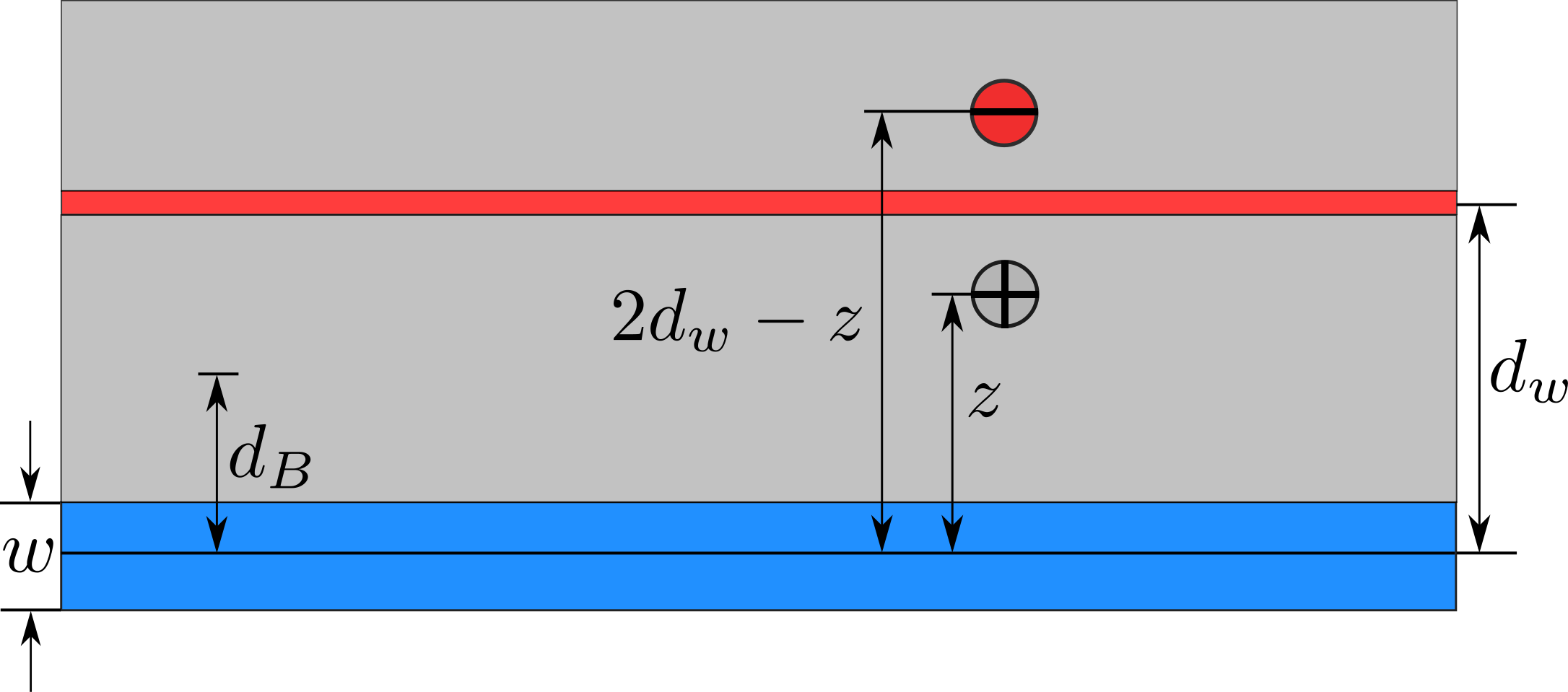}
 	\caption{The upper half of the structure shown in Fig.\,1(a) with an impurity (plus) at a distance $z$ from the midplane of the 2DEG. Due to EES, an image charge (red minus) is produced at a distance $2d_w-z$ from the midplane that reduces the potential of the impurity. }
 	\label{fig:image}
 \end{figure}

Let us now show that EES truncates the divergence even stronger. We assume that $f\gtrsim f_c$ so that EES is already so strong that its screening radius $r_s\ll d_w$, i.e. the doping layer screens a static potential as if it were a metal. 
This means that at $f\geq f_c$, $\mu_B$ and $\mu_{q,B}$ are independent of $f$. Let us consider an impurity at a distance $z>0$ from the midplane of the 2DEG, as shown in Fig.\,\ref{fig:image}.
When the impurity is located between the 2DEG and the doping layer, the doping layer creates an oppositely charged image of the impurity at the distance $2d_w - z$ which reduces its potential. When $z\ll d_w$, the image is far from the 2DEG and its effect is small. 
However, when $z$ becomes close to $d_w$ as shown in Fig.\,\ref{fig:image} the impurity forms a compact dipole with its image whose potential in the 2DEG practically does not scatter. 
Thus, the scattering off of impurities located at distances larger than a critical distance $z=d_B$ becomes negligible.
We can estimate $d_B$ by solving the equation
 \begin{equation}
	\left(\frac{e}{\kappa d_B}-\frac{e}{\kappa(2d_w-d_B)}\right)^2=\frac{1}{2}\left(\frac{e}{\kappa d_B}\right)^2\,,
 \end{equation}
which gives $d_B \simeq 0.5d_w$. Here we use the squares of the potentials as they lead to scattering. Furthermore, we can ignore the impurities at $z > d_w$, as the doping layer acts as a Faraday cage which screens these impurities.\cite{note:3}
Thus we need to consider only the impurities within the finite distance $\abs{z}< d_B$. 

Now let us make our model more general and assume that the BI concentrations outside and inside the well are $N_1$ and $N_2$, respectively. 
Then we can write the linear Eqs.\,(\ref{eq:mobility_linearnumbers}) and (\ref{eq:quantummobility_linearnumbers}) for the mobilities $\mu_B$ and $\mu_{q,B}$ with coefficients $A_1$, $A_2$, $B_1$, and $B_2$ which are estimated below. 
 
Let us first concentrate on $A_1$ and $B_1$ related to the impurities in the Al$_x$Ga$_{1-x}$As barriers. Due to EES, only the impurities of the layer $w/2<\abs{z}<d_B$ contribute to scattering. Because $a_B/2, w/4 < w/2\ll d_B$, for the purpose of estimates we can apply Eqs.\,(\ref{eq:mobilityAndo_approx}) and (\ref{eq:quantummobilityAndo_approx}) to a thin layer of impurities between $z$ and $z+dz$ with concentration $N_1dz$, and sum the contributions of these layers arriving at 
\begin{align}\label{eq:mobility_remotenoEES}
\mu^{-1}_B= \frac{\pi}{4}\frac{\hbar}{e} \int\limits_{w/2}^{d_B} \frac{N_1dz}{(k_{F} z)^3} =A_1N_1\,,\\
\mu_{q,B}^{-1}= \pi\frac{\hbar}{e} \int\limits_{w/2}^{d_B} \frac{N_1dz}{k_Fz}=B_1N_1\,, \label{eq:quantummobility_remotenoEES}
\end{align} 
where we multiplied by 2 as these impurities lie on both sides of the 2DEG. We find\cite{note:4}
 \begin{align}\label{eq:A_1}
 A_1\simeq\frac{\pi}{2}\frac{\hbar}{e}\frac{1}{k_F^3w^2}\,,\\
 B_1\simeq\pi\frac{\hbar}{e}\frac{1}{k_F}\ln\left(\frac{d_w}{w}\right)\,. \label{eq:B_1}
\end{align} 

Let us switch to $A_2$ and $B_2$ which are determined by impurities in the GaAs well. 
Here, in order to get a very rough estimates we do not discriminate between the two smallest spatial scales, the screening radius of 2DEG $a_B/2 \simeq 5$ nm, and the ``half width of electron layer" $w/4 = 7.5$ nm. Then we can write

\begin{align}\label{eq:A_2}
A_2 \approx \frac \pi 4 \frac{\hbar}{ ek_F} \left(1 + k_F\int\limits_{w/4}^{w/2} \frac{dz}{(k_F z)^3}
% + \int\limits_{0}^{w/4} \frac{dz}{(k_F w/4)^3} 
\right ) \simeq 17 \frac{\hbar}{e}\frac{1}{k_F^3 w^2}\,,\\
B_2 \approx \pi \frac{\hbar}{ek_F} \left( 1 + k_F\int\limits_{w/4}^{w/2} \frac{dz}{k_F z} 
%+ \int\limits_{0}^{w/4} \frac{dz}{k_F w/4} 
\right) \simeq 5\frac \hbar e \frac 1 {k_F}\,,\label{eq:B_2}
\end{align} 
where we took into account that $k_F w/4 \simeq 1$ and replaced $z$ by $w/4$ when integrating from $0$ to $w/4$.

Using the reference sample parameters,\cite{note:pars}, and expressing the mobilities in units $1\times10^{6}$ cm$^{2}$V$^{-1}$s$^{-1}$ and $N_1$ and $N_2$ in units of $10^{14}$ cm$^{-3}$, we find that Eqs.\,(\ref{eq:A_1}), (\ref{eq:B_1}), (\ref{eq:A_2}), (\ref{eq:B_2}) give $A_1=0.005$, $B_1=0.20$, $A_2=0.06$ and $B_2=0.25$.
In the Appendix we develop a quantitative theory of BI scattering which 
%using form factors of the quantum well and using the RPA approximation. This theory 
confirms the estimates $A_1=0.005$, $B_1=0.20$ and leads to slightly larger $A_2=0.10$ and $B_2=0.33$.

Now one can easily calculate $N_1$ and $N_2$ from $\mu_B$ and $\mu_{q,B}$ solving Eqs.\,(\ref{eq:mobility_linearnumbers}) and (\ref{eq:quantummobility_linearnumbers}). For example, if $\mu=3\times10^7$ cm$^2$V$^{-1}$s$^{-1}$ and $\mu_q=1\times10^6$ cm$^2$V$^{-1}$s$^{-1}$, at $f=f_c$ we can subtract the contribution of $\mu_{R}$ and $\mu_{q,R}$ from $\mu$ and $\mu_q$ and find $\mu_B = 33 \times 10^6$ cm$^2$V$^{-1}$s$^{-1}$ and $\mu_{q,B} = 2 \times 10^6$ cm$^2$V$^{-1}$s$^{-1}$. They correspond to $N_1 \simeq 2\times 10^{14}$ cm$^{-3}$ and $N_2\simeq2\times10^{13}$ cm$^{-3}$. In this case Al$_x$Ga$_{1-x}$As accounts for nearly $90\%$ of $\mu_{q,B}$, while it accounts for only $40\%$ of $\mu_B$. This large sensitivity to the Al$_x$Ga$_{1-x}$As impurities implies that for improvements one should focus not only on the Ga purity,\cite{gardner:2016} but also on the Al purity as well, particularly if one is interested in high $\mu_{q,B}$.

Previous work\cite{Macleod} has shown that the ratio $\mu_B/\mu_{q,B}$ without EES in single heterojunction devices is $\sim 10$. As EES reduces BI scattering, one might expect that this ratio would decrease. Yet in the above discussion, we have shown that the ratio $\mu_B/\mu_{q,B}$ can be as large as 18. This large number is a result of allowing $N_1$ and $N_2$ to be different, while Ref.\,[\onlinecite{Macleod}] assumed that $N_1=N_2$. The large ratio $N_1/N_2$ compensates the reduction in BI scattering by EES. If we assume that $N_1=N_2$, then it follows from Eqs.\,(\ref{eq:mobility_linearnumbers}) and (\ref{eq:quantummobility_linearnumbers}) with $A_1=0.005$, $B_1=0.20$, $A_2=0.10$ and $B_2=0.33$ that EES reduces the ratio to $\mu_B/\mu_{q,B}\simeq5$. 

Let us now consider a heterostructure in which the bottom doping layer is removed to allow tuning of the electron concentration in the 2DEG by a back gate placed at a distance $L \simeq 800$ nm below the 2DEG.\cite{qian:2017b,Zudov_widewell} 
In this case we expect the scattering from the impurities below the 2DEG to increase as distant impurities, which were previously screened by the excess electrons, now contribute to scattering. 

%We do not expect this to affect $\mu_B$, as these distant impurities scatter at small angles. However, $\mu_{q,B}$ is sensitive to distant impurities and the coefficient $B_1$ [see Eq.\,(\ref{eq:B_1})] needs to be modified.

The gate also produces images of background impurities which screen their static potential (see Fig. 4).
Therefore, we can modify the parameter $B_1$ by replacing the EES screening length $d_B$ by the gate screening length $L/2$ for the bottom layer and write
\begin{equation}\label{eq:A(L)}
B_1(L)\simeq\frac{B_1}{2}\left[1+\frac{\ln(L/w)}{\ln(2d_B/w)}\right]\,.
\end{equation}
For the reference sample parameters\citep{note:pars} and $L=800$ nm we find $B_1(L)\simeq 1.9 B_1$. 
This finding seems to be in qualitative agreement with experiments, which reported higher $\mu_q$ in heterostructures with two SPSL doping layers\cite{shi:2016a,shi:2017a} than in a gated heterostructure\cite{qian:2017b} of the same $w$ and tuned to the same $n_e$. 
Note, however, that the difference in $\mu_q$ could also originate from other factors, such as different $f$ or $N_1$.

Above we have assumed that the Born approximation is valid everywhere and that the logarithmic divergence is truncated by the EES of impurities. In principle, one may go beyond the Born approximation and use a self-consistent multiple scattering theory to truncate logarithmic divergence,$^{47,60}$ 
which introduces a truncation length on the order of $k_F^2/N_1$. 
For modern samples, $N_1\sim10^{14}$ cm$^{-3}$ and the distance $k_F^2/N_1\sim 0.2$ mm, which is significantly larger than either $d_w/2$ or $L/2$, so that there is no need to use the self-consistent multiple scattering theory.

%Furthermore, we can find from Eqs.\,(\ref{eq:B_1}) and (\ref{eq:A(L)}) that $\mu_{q,B}\propto k_F=(2\pi n_e)^{1/2}$, so we expect that $\mu_{q,B}$ should increase with density. 

\section{Additional disorder in \\ doping layers}\label{sec:disorder}

So far we have considered only two sources of disorder, namely, the random location of donors in a single layer in the middle of the GaAs doping well and the random three-dimensional distribution of impurities with concentrations $N_1$ and $N_2$ in the Al$_x$Ga$_{1-x}$As and the GaAs, respectively. 
In this idealized model, the screening radius decreases very fast with $f$, making $\mu_{q,R}$ quickly approach and exceed $\mu_{q,B}$.

In this section we discuss three additional sources of disorder in the doping layer, which can reduce the density of states of excess electrons, increasing their screening radius and, thus, suppressing EES. 
Therefore, our Eqs.\,(\ref{eq:mobility_final}), (\ref{eq:quantummobility_final}) and Eqs.\,(\ref{eq:mobility_linearnumbers}), (\ref{eq:quantummobility_linearnumbers}) are the best mobilities achievable in the SPSL-doped structures. 
%Below we discuss three additional sources of disorder.

We start from the possible spreading of the Si donors over several layers of the GaAs well. 
The variations $\delta s$ in the distance $s$ between donors and the middle plane of the AlAs layer leads to variations in the binding energy $E_b$ of dipole atoms, reducing the density of states of the excess electrons. 
Assuming that $\delta s = a$, where $a \simeq 0.5$ nm is the lattice constant, and using Fig.\,\ref{fig:variation} we can estimate the correction to the binding energy $\delta E_b \simeq 0.2$ $Ry \simeq 5$ meV. 
Since this correction is smaller than the typical Coulomb width of the dipole density of states $E_C=e^2n^{1/2}/\bar{\kappa}\simeq12$ meV, it should only lead to an increase of the screening radius $r_s$ by a relatively small fraction $\sim(\delta E_b/E_C)^2$. 
If, on the other hand, Si donors are spread over larger $\delta s$, the RI-limited mobilities might degrade. 
%However, in the range $0.15< f< 0.35$ the $f$-dependence of $E_F$ leading to Eq.\,(\ref{eq:screeningradiusnumerical}) is rather stable with respect to random energy shifts\cite{Bello} and this effect is expected to be small until $f$ becomes larger.

Another source of disorder is the roughness of AlAs/GaAs and AlAs/Al$_x$Ga$_{1-x}$As interfaces.\cite{Sakaki,Hanroughness,Gold87}
If we assume that these interfaces are covered by random islands of height $a$ and radius $R$ larger than the localization length $\xi$, but smaller than the distance $\nd^{-1/2}$ between donors, the dispersion $\Gamma$ of the first subband quantization energies of the excess electrons can be estimated as
\begin{equation}\label{eq:roughness}
\Gamma = \frac{2a}{t} \frac{\hbar^{2}\pi^2}{2m_z^\star t^{2}} \simeq 45 \mbox{ meV}\,.
\end{equation}
If we assume that the density of states of dipole atoms with energy $E$ is dominated by roughness and has a Gaussian shape $g(E)=n\pi^{-1/2}\Gamma^{-1} \exp[- (E/\Gamma)^2]$, then we can find the Fermi energy $E_F (f)$ and arrive at the estimate of the screening radius for small $f$ 
\begin{equation}\label{eq:screeningradius_gamma}
r_s(f) \simeq \frac{\bar{\kappa}\Gamma}{4\pi e^2 f\nd}\,.
\end{equation}
Comparing Eqs.\,(\ref{eq:screeningradiusnumerical}) and (\ref{eq:screeningradius_gamma}), we find that they cross at $f\simeq 0.36 \simeq f_c$, so that roughness only weakly perturbs the crossover between RI-limited and BI-limited $\mu_q$ for the reference sample parameters.\cite{note:pars} However, if BI scattering is reduced by making the Al$_x$Ga$_{1-x}$As barriers cleaner, then roughness will be the main limiting factor of $\mu_q$ at large $f$.
Since the roughness parameters are not well known, we do not attempt to make more accurate estimates.

The third source of additional disorder comes from possible autocompensation of Si donors. 
If a small fraction $\alpha$ of Si atoms resides in the As sublattice, they are acceptors. 
A typical acceptor traps an electron from the nearest neighbor donor forming an acceptor-donor dipole with the arm of the order of $\nd^{-1/2}$ in the $x-y$ plane. 
If we still use $\nd$ for the concentration of Si atoms and if the concentration of holes forming the Wigner crystal on top of them $(1-f)\nd$ is larger than the concentration of acceptor-donor dipoles $\alpha \nd$, these dipoles produce a minor effect on the random potential. 
Only at $(1-f) <\alpha$ do the randomly oriented acceptor-donor dipoles dominate the random potential making it $f$-independent. 
If, for an estimate, we follow Ref.\,[\onlinecite{Bose}] which found $\alpha \sim 0.1$, this source of disorder becomes relevant only when $f$ approaches unity and thus has minor effect on our findings.

%As we saw in this section our results for $0.15 < f < 0.35$ are relatively robust against either roughness of GaAs/AlAs interfaces or donor spreading in GaAs doping well. 
%On the other hand, these effects become very important for $1-f \ll 1$ where they aggressively destroy the Wigner crystal of holes mentioned at the end of Sec.\,\ref{sec:remote}. This implies that $\mu_R$ and $\mu_{q,R}$ likely saturate at some $f \geq 0.5$ due to roughness and/or donor spreading.

\section{Quantum Hall effect at $\nu = 5/2$}\label{sec:5/2QHE}

In this section we would like to comment on the puzzle of the experimentally obtained gap $\Delta_{5/2}^{exp}$ of the quantum Hall effect at filling factor $\nu=5/2$.\cite{ManfraReview,nuebler:2010,Morf}
The observed $\Delta_{5/2}^{exp} \lesssim 0.7$ K is considerably smaller than the theoretical value of $\Delta_{5/2}^{th} \simeq 2$ K and we would like to see if our theory can shed the light on this issue.

If in the absence of 2DEG screening the magnitude of the long range fluctuations of the potential energy of 5/2 excitations $V \lesssim \Delta_{5/2}^{th}$, they should not affect $\Delta_{5/2}^{exp}$ as it is determined by the classical trajectories of excitations with activation over the saddle points.\cite{Polyakov1995} 
However, when $V \gg \Delta_{5/2}^{th}$ the disorder is nonlinearly screened by the 2DEG and creates large compressible islands separated by relatively narrow stripes of incompressible liquid.\cite{Efros1988,Pikus1993} 
In this case, $\Delta_{5/2}^{exp}$ can be substantially smaller than $\Delta_{5/2}^{th}$ due to the following effects.\cite{Morf} 
First, the self-energy of charged excitations created in the incompressible stripes is reduced by the proximity of the excitations to the metal-like compressible islands. 
Second, the tunneling through the saddle points of $V$ can now happen at smaller distances which are comparable with the size of the excitation. 
Here, we would like to show that at $f=f_c$ the disorder is already so weak due to EES that $V \sim \Delta_{5/2}^{th}$ and thus there is no reason to expect a substantial deficit of $\Delta_{5/2}^{exp}$.

Even though at $f=f_c$ the remote donors and background impurities provide equal contributions to $\mu_q^{-1}$, their effective concentrations, which determine the spatial scales of their random potentials, are different. 
Indeed, comparing Eqs.\,(\ref{eq:quantummobility_final}) and (\ref{eq:quantummobilityAndo_approx}) we can find that at $f=f_c$ EES reduces $\nd$ to an effective concentration $n_s$ of randomly positioned charged donors 
%\begin{equation}\label{eq:effectiveconcentration}
$n_s \simeq 0.57/d_w^{2} \approx 6\times 10^{9}\ \text{cm}^{-2} \ll \nd.$
%\end{equation}
On the other hand, the BI potential is due to impurities in the layers of width $0.5d_w$ located on both sides of the 2DEG. 
For $N_1 \simeq 2\times10^{14}$ cm$^{-3}$, the two-dimensional concentration of such impurities is $n_B = d_w N_1 \approx 2.1\times 10^9$ cm$^{-2} \approx n_s/3$. 
As a result, the spatial scale $n_B^{-1/2}$ of the BI potential is larger than the scale $n_s^{-1/2}$ of the RI potential.\cite{note:6}

Due to their smaller spatial scale $n_s^{-1/2}$, the random fluctuations of the remote donor potential $\sim en_s^{1/2}/\kappa$ are responsible for tunneling at a saddle point. 
For fluctuations of the potential energy of the 5/2 excitations with charge $e/4$ this translates to $V\sim e^{2}n_{s}^{1/2}/4\kappa \approx 2~{\rm K}\simeq\Delta_{5/2}^{th}$.  
Thus the compressible islands of the 2DEG should be small and play marginal role. 
Then the sum of the two self-energies necessary to create two oppositely-charged excitations should be close to $\Delta_{5/2}^{th}$ and the characteristic tunneling distance at the saddle point of $V$ should be $n_s^{-1/2}$. 
This distance should be compared to the size of the charge $e/4$ excitations $2 l_B$,\cite{Morf} where $l_B = (\hbar c/eB)^{1/2}$ is the magnetic length. 
At $B=5$ T, $2l_B\simeq 23$ nm and $n_s^{-1/2}$ is five times larger than $2 l_B$.
Thus we expect that tunneling through saddle points plays a very weak role and there should be a range of temperatures in which the transport is activated with no deficit.\cite{Polyakov1995} 
Of course, at very low temperatures one should expect that tunneling eventually becomes important and hopping transport takes over. 
At $f < f_c$, the EES is weaker so that the amplitude of the potential energy fluctuations $V$ is larger and leads to large compressible islands and narrow incompressible strips which can substantially reduce $\Delta_{5/2}^{exp}$.\cite{umansky:2009}

Above we assumed that the random positions of donors is the only source of disorder. As shown in the previous section, additional sources of disorder can weaken the effects of EES and lead to the deficit of $\Delta_{5/2}^{exp}$. 
It is therefore plausible that reduction of these sources of disorder can lead to the increase of $\Delta_{5/2}^{exp}$. 

\section{Summary}\label{sec:summary}

We have studied the mobility and quantum mobility in GaAs/Al$_x$Ga$_{1-x}$As heterostructures with SPSL-doping (see Fig.\,1). 
We showed that scattering by both remote donors and charged background impurities is strongly reduced due to screening of their random potentials by excess electrons. 
To evaluate the strength of this screening, we considered a dipole atom formed by a donor and an excess electron in the SPSL layer and showed that excess electrons are localized and have negligible parallel-to-2DEG conductance. 
On the other hand, excess electrons strongly screen the random potential of the donors. 
We calculated the screening radius of the excess electrons and used it to calculate the remote donor-limited mobilities $\mu_R$ and $\mu_{q,R}$ [Eqs.\,(\ref{eq:mobility_final}) and (\ref{eq:quantummobility_final})]. 
Our estimates show that there is a characteristic filling fraction $f_c$ of excess electrons beyond which not only $\mu_R$ but also $\mu_{q,R}$ exceeds the background impurity-limited mobilities $\mu_B$ and $\mu_{q,B}$, respectively.
Furthermore, the screening by excess electrons is so strong that $\mu_B$ and $\mu_{q,B}$ are determined only by impurities which are located within a distance $0.5d_w$ from the midplane of the 2DEG. Our calculations of $\mu_B^{-1}$ and $\mu_{q,B}^{-1}$ as linear functions of the concentrations of impurities in Al$_x$Ga$_{1-x}$As and in GaAs [Eqs.\,(\ref{eq:mobility_linearnumbers}) and (\ref{eq:quantummobility_linearnumbers})] provide a way to estimate these concentrations if $\mu_B$ and $\mu_{q,B}$ are known. 
We generalized our results for background impurity scattering to backgated heterostructures, in which the bottom doping layer is removed, and found that in these heterostructures $\mu_{q,B}$ should be roughly twice smaller than in heterostructures with two doping layers.
We discussed additional disorder in the doping layers due to the spreading of the $\delta$-layer of donors in the GaAs doping well and due to roughness of the AlAs/GaAs and AlAs/Al$_x$Ga$_{1-x}$As intefaces. 

%Both of them might need to be minimized in order to maximize $\mu_q$ and improve quality of the 5/2 quantum Hall effect.

In conclusion, we would like to summarize possible avenues for improvement of the quantum mobility in modern GaAs/Al$_x$Ga$_{1-x}$As heterostructures. To increase $\mu_{q,R}$, one should ensure that $f>f_c$ in both doping layers and, if necessary, find ways to minimize the spreading of donors and the roughness of the AlAs/GaAs and AlAs/Al$_x$Ga$_{1-x}$As interfaces. To increase $\mu_{q,B}$, one should find a way to make the Al$_x$Ga$_{1-x}$As barriers cleaner, as the impurities in these layers are what limits $\mu_{q,B}$. 

%In conclusion, we would like to repeat our suggestions for improvement of the quantum mobility of modern GaAs heterostructures. 
%First, anything should be done to increase the filling fraction $f$ of the least populated doping layer beyond $f_c$.  
%However, while this is necessary, it may not be sufficient because even for large enough $f$ the excess electron screening can be suppressed by additional disorder in the doping layers.
%Second, one should reduce the number of impurities in the Al$_x$Ga$_{1-x}$As barriers as we found that these impurities are responsible for nearly all of the contribution from background impurities to the quantum mobility.

$\phantom{}$
\vspace*{2ex} \par \noindent
\begin{acknowledgments}
We are grateful to M. J. Manfra, L. N. Pfeiffer, and V. Umansky for discussions of growth and sample details, and Q. Shi, H. Fu, and X. Ying for their commenting of the manuscript and helpful discussions. M. Sammon was supported primarily by the NSF through the University of Minnesota MRSEC under Award No. DMR-1420013. 
M.A.Z. acknowledges support by the U.S. Department of Energy, Office of Science, Basic Energy Sciences, under Award No. ER 46640-SC0002567 and by the NSF Award No. DMR-1309578.
\end{acknowledgments}
\vspace*{3ex}

\appendix

\section*{Appendix:Quantitative Theory of Scattering by Background Impurities}
\renewcommand{\theequation}{A.\arabic{equation}}
\setcounter{equation}{0}
We begin by deriving the screened interaction between a 2DEG in a quantum well of width $w$ and an impurity at a distance $z$ away from the center of a quantum well ($z=0$). 
The impurity is screened by two electron gases $-$ the 2DEG inside of the quantum well and the excess electrons (EE) in the nearest doping layer a distance $|z| = d_w$ away from the center of the main quantum well (See Fig.\,\subref{fig:device}). 
We ignore the thickness of the doping layer (as it is small compared to $d_w$) and consider the screening by only the doping layer nearest to the impurity. 
 
The screening of a single impurity is calculated using the random phase approximation (RPA), which amounts to writing a set of self consistent equations for the screened interactions between a charged impurity $i$ and the electrons. 
Here, we introduce the notation $U_{i,k}$ for the screened impurity interactions and $\widetilde{U}_{i,k}$ for the bare impurity interactions, where the subscript $k$ can be 1 (2DEG) or 2 (EE). 
Screening occurs because of the electron-electron interactions $\widetilde{U}_{k,l}$, which can be 2DEG-2DEG ($k=l=1$), EE-EE ($k=l=2$) or 2DEG-EE ($k=1,l=2$). 
We assume that the electrons are completely confined inside the well and occupy only the first subband, so their linear density in the $z$-direction is $\lambda_1(z)=(2/w)\cos^2(\pi z/w)\Theta(w/2-|z|)$. 
Conversely, since we ignore the thickness of the doping layer, the linear density of the EE is given by $\lambda_2(z)=\delta(z-d_w)$. 
Using this notation, the RPA gives the system of equations 
\begin{align}\label{eq:RPA}
U_{i,k}=\widetilde{U}_{i,k}+\sum\limits_{l=1}^{2}U_{i,l}\Pi_l\widetilde{U}_{l,k}\,,
\end{align}
where $\Pi_k$ are polarization functions of the electrons,
\begin{equation}\label{eq:I_el}
\widetilde{U}_{i,k}=\left(\frac{2\pi e^2}{\kappa q}\right)\int dz' \lambda_k(z')e^{-q|z-z'|}
\end{equation}
are the bare impurity-electron interactions, and
\begin{equation}\label{eq:el_el}
\widetilde{U}_{k,l}=\left(\frac{2\pi e^2}{\kappa q}\right)\int dz \int dz'\lambda_k(z)\lambda_l(z')e^{-q|z-z'|}
\end{equation}
are the electron-electron interactions. 
Solving for $U_{i,k}$, we find 
\begin{equation}\label{eq:RPA_sol}
U_{i,1}=\frac{\widetilde{U}_{i,1}(1-\Pi_2\widetilde{U}_{2,2})+\widetilde{U}_{i,2}\Pi_2\widetilde{U}_{2,1}}{(1-\Pi_1\widetilde{U}_{1,1})(1-\Pi_2\widetilde{U}_{2,2})-\Pi_1\Pi_2\widetilde{U}_{1,2}^2}\,.
\end{equation} 
\begin{widetext}
The bare impurity interactions are straightforward to calculate from Eqs.\,(\ref{eq:I_el}) and (\ref{eq:el_el}). For $\widetilde{U}_{i,1}$ we find
\begin{equation}\label{eq:U_{I,1}}
\widetilde{U}_{i,1}(q,z)=\left(\frac{2\pi e^2}{\kappa q}\right)F_0(qw)\begin{cases}
\mbox{csch}\left(\frac{qw}{2}\right)\left[1-\exp\left(\frac{-qw}{2}\right)\cosh(qz)\right], & \abs{z}<w/2\\
\exp(-qz), & \abs{z}>w/2\,,
\end{cases}
\end{equation}
\end{widetext}
where 
\begin{equation}
F_0(x)=\frac{8\pi^2}{x[x^2+4\pi^2]}\sinh\left(\frac{x}{2}\right)\,.
\end{equation}
For $\widetilde U_{i,2}$ we find
\begin{equation}\label{eq:U_{I,2}}
\widetilde{U}_{i,2}(q,z)=\left(\frac{2\pi e^2}{\kappa q}\right)\exp(-q|z-d_w|).
\end{equation}
Similarly, we find for the electron-electron interactions that
\begin{align}\label{eq:U_{1,1}}
&\widetilde{U}_{1,1}=\left(\frac{2\pi e^2}{\kappa q}\right)G(qw)\,,\\
&\widetilde{U}_{1,2}=\left(\frac{2\pi e^2}{\kappa q}\right)e^{-qd_w}F_0(qw)\,,\label{eq:U_{1,2}}\\
&\widetilde{U}_{2,2}=\left(\frac{2\pi e^2}{\kappa q}\right)\,,\label{eq:U_{2,2}}
\end{align}
where
\begin{equation}
G(x)=\frac{20\pi^2x^3+3x^5-32\pi^4(1-e^{-x}-x)}{x^2(4\pi^2+x^2)^2}\,.
\end{equation}
For the polarization functions we use the Thomas-Fermi approximation which gives $\Pi_1=-\kappa q_{TF}/2\pi e^2$ and $\Pi_2=-\kappa r_s^{-1}/2\pi e^2$.

Let us now examine Eq.\,(\ref{eq:RPA_sol}) at different $z$. 
For simplicity, we will set $w=0$ and assume $2qd_w\ll 1$, as small $q$ give the main contribution to $\mu_{q,B}$. 
For $z \ll d_w$, Eq.\,(\ref{eq:RPA_sol}) with the interactions given in Eqs.\,(\ref{eq:U_{I,1}})-(\ref{eq:U_{2,2}}) reduces to 
\begin{equation}
U_{i,1}(q,z)\simeq \frac{2\pi e^2}{\kappa q_{TF}}\,,
\end{equation}
so that an impurity at small $z$ is screened by the 2DEG but not by the EE. 
Conversely, for $z \gtrsim d_w$ we find
\begin{equation}
U_{i,1}(q,z) \simeq\frac{r_se^{-qz}}{2d_w}\left(\frac{2\pi e^2}{\kappa q_{TF}}\right)\,,
\end{equation} 
so that at larger $z$ $U_{i,1}$ is suppressed by a factor $2d_w/r_s$. 
For $f=f_c$ and reference sample parameters,\cite{note:pars} $2d_w/r_s \simeq 23$ and the impurities at $z \gtrsim d_w$ play no role.

%\vspace{2ex}

The above discussion allows us to make two assumptions that substantially simplify the calculations. 
First, we ignore EES for the impurities in the GaAs well ($\abs{z}<w/2$). 
Second, we ignore the impurities in the Al$_{x}$Ga$_{1-x}$As beyond the doping layer$^{54}$ ($\abs{z}>d_w$). 
Under these assumptions, $\mu_B$ and $\mu_{q,B}$ can be calculated using the Born approximation according to Eqs.\,(\ref{eq:mobilityBG}) and (\ref{eq:quantummobilityBG}). The scattering potential $\la|U(q)|^2\ra$ is given by
\begin{equation}\label{eq:potential_general}
\la|U(q)|^2\ra=\int\limits_{-\infty}^{\infty}N(z)U_{i,1}^{2}(q,z) dz\,,
\end{equation} 
where 
$$
N(z)=\begin{cases}
N_1\,, &\abs{z}>w/2\\
N_2\,,&\abs{z}<w/2
\end{cases}
$$
is the 3D concentration of impurities at a distance $z$ from the center of the 2DEG, $N_1$ ($N_2$) is the concentration of impurities in Al$_x$Ga$_{1-x}$As (GaAs), and $U_{i,1}(q,z)$ is the impurity-electron interaction with EES defined in Eq.\,(\ref{eq:U_{I,1}}). Performing the integration in Eq.\,(\ref{eq:potential_general}) yields
\begin{widetext}
	\begin{equation}\label{eq:potential_}
	\la|U(q)|^2\ra= \left(\frac{2\pi e^2}{\kappa q}\right)^2\left[\frac{N_1}{q}\frac{F_1(qw,qd,qr_s)}{\varepsilon^2(q)}
	+\frac{N_2}{q}\frac{F_2(qw)}{\varepsilon_0^2(q)}\right]\,,
	\end{equation}
	where 
	\begin{equation}\label{eq:formfactor_AlGaAs}
	F_1(x,y,z)=e^{-x}F_0^2(x)\left[\frac{(1+z)^2-e^{-2y}[z(2+z)+4y(z+1)]-e^{-4y}}{z^2}\right],
	\end{equation}
	and 
		\begin{equation}\label{eq:formfactor_GaAs}
		F_2(x) = 
		\frac{1}{x}\left(\frac{4\pi^2}{4\pi^2+x^2}\right)^2\left[\frac{8e^{-x}-e^{-2x}-7}{x}+
		2(2+e^{-x})+\frac{2x^2}{\pi^2}+\frac{3x^4}{8\pi^4}-\frac{8x(1-e^{-x})}{4\pi^2+x^2}\right],
		\end{equation}
		are the form factors\cite{Gold87} for Al$_{x}$Ga$_{1-x}$As and GaAs respectively, while 
\begin{equation}\label{eq:screening_function}
\varepsilon(q)=\left(1+G(qw)\frac{q_{TF}}{q}\right)\left(1+\frac{1}{qr_s}\right)-F_0^2(qw)\frac{q_{TF}}{q^2r_s} e^{-2qd_w}
\end{equation}
and
\begin{equation}\label{eq:screening_2DEG}
\varepsilon_0(q)=1+G(qw)\frac{q_{TF}}q
\end{equation}
are the dielectric functions with and without EES. 
\end{widetext}
With Eqs.\,(\ref{eq:potential_})$-$(\ref{eq:screening_2DEG}), we can find the coefficients $A_1$, $A_2$, $B_1$, and $B_2$ in Eqs.\,(\ref{eq:mobility_linearnumbers}) and (\ref{eq:quantummobility_linearnumbers}). 
Using the reference sample parameters\citep{note:pars} we find $A_1=0.005$, $A_2=0.10$, $B_1=0.20$, and $B_2=0.33$.


\begin{thebibliography}{75}
\bibitem[{not({\natexlab{e}})}]{note:2}
	\bibinfo{note}{At small enough $f$, when the density of states of the SPSL
		structure is small, localization can be achieved even at
		$n\xi^2\simeq(n\xi^2)_c$.}
\bibitem[{not({\natexlab{f}})}]{note:9}
	\bibinfo{note}{The reason is that at $f\geq0.15$ the correction to Eq.
		(\ref{eq:EF_1-K}) from the long-range potential fluctuations becomes
		substantial.}
\bibitem[{not({\natexlab{h}})}]{note:3}
	\bibinfo{note}{For these impurities the scattering potential is reduced by the
		factor $4(d_w/r_s)^2\sim 520$ at $f=f_c$. Strictly speaking the logarithmic
		divergence of $\mu_{q,B}^{-1}$ exists even in this case. However, it is
		substantial only for the sample thickness $L\sim\exp(4(d_w/r_s)^2)$ which is
		unrealistically large and so we can safely ignore the impurities located
		behind the doping layers.}
\bibitem[{not({\natexlab{j}})}]{note:6}
	\bibinfo{note}{While $n_s \approx 3n_B$ at $f=f_c$, the BI and RI contribute
		equally to $\mu_q^{-1}$. The reason for this is that $\mu_q^{-1}\propto\int
		d\theta \sigma(\theta)$, where $\sigma(\theta)$ is the scattering
		cross-section. The cross section $\sigma(\theta) \sim \exp(-k_Fz\theta)$
		decays quickly with $z$, so that the larger it is, the smaller the range of
		$\theta$ that contributes to scattering.}
\end{thebibliography}
\end{document}